\documentstyle[preprint,aps]{revtex}

\begin{document}
\preprint{UNDPDK-98-01}
\title{Measuring the $\nu_{\mu}$ to $\overline{\nu_{\mu}}$ Ratio in a
High Statistics Atmospheric Neutrino Experiment}
\author{J.M.~LoSecco}
\address{University of Notre Dame, Notre Dame, Indiana 46556}
\date{June 3, 1998}
\maketitle
\widetext
\begin{abstract}
By exploiting differences in muon lifetimes it is possible to distinguish
$\nu_{\mu}$ from $\overline{\nu_{\mu}}$ charged current interactions in
underground neutrino detectors.  Such observations would be a useful tool
in understanding the source of the atmospheric neutrino anomaly.\\
Subject headings: Cosmic Rays --- Elementary Particles --- Neutrino Oscillations\\
\end{abstract}

\section{Introduction}
The atmospheric neutrino anomaly \cite{haines,kamioka,imbo} is the discrepancy
between the observed and expected rate of electron and muon neutrino
interactions in underground detectors.  In general it is believed that
these neutrinos originate in the Earth's atmosphere as a consequence of
the decay of short lived particles created by cosmic ray interactions.

The most popular explanation for the source of the anomaly is the oscillation
of muon neutrinos.
A number of hypothetical solutions have suggested a new
form of interaction.  For example Ma and Roy \cite{Ma} point out that
a new diagonal neutral current interaction for the $\nu_{\tau}$ could
produce a coherent picture for all current neutrino problems
(solar\cite{solar},
atmospheric and LSND\cite{lsnd}).  Such new interactions would have
different effects on neutrinos and antineutrinos so there is some interest
in distinguishing $\nu$ from $\overline{\nu}$.  In general a charged current
neutrino interaction produces a charged muon or electron.  The sign of the
charge can be used to infer the particle/antiparticle nature of the
interacting neutrino.

Morphological methods have been employed to distinguish charged current
muon and electron events.  But the effect was initially recognized when the
fraction of event containing a muon decay signature was considerably below
expectations\cite{haines}.

This paper points out that CPT violating differences in the detector
itself make it possible to distinguish on a statistical basis between
$\nu_{\mu}$ and $\overline{\nu_{\mu}}$ induced interactions.

\section{The Method}
Due to the possibility of muon capture \cite{muonO} the
$\mu^{-}$ has a larger decay
width than $\mu^{+}$ when stopped in normal matter.
\[
\Gamma_{\mu^{+}} = \Gamma_{\mu} = 1/\tau_{+}
\]
\[
\Gamma_{\mu^{-}} = \Gamma_{\mu} +  \Gamma_{Capture} = 1/\tau_{-}
\]

This leads to a shorter lifetime for $\mu^{-}$ than for $\mu^{+}$ when they
decay in ordinary matter.  The effect may not be large.  It is about 18\%
for muons in water but increases with $Z$ so is more pronounced in heavier
materials.

The observed time distribution for muon decays is the weighted sum of the
two exponential decay distributions.

\[
f_{-} e^{-t/\tau_{-}} + (1-f_{-}) e^{-t/\tau_{+}}
\]
where $f_{-}$ is the fraction of decays due to a $\mu^{-}$ and
$\tau_{-}$ and $\tau_{+}$ are the {\em known} decay lifetimes for the
$\mu^{-}$ and $\mu^{+}$ respectively in the detector environment.

The mean value of the measured lifetime of a mixture of $\mu^{-}$ and $\mu^{+}$
is then:
\[
\tau_{Observed} = f_{-} \tau_{-} + (1-f_{-}) \tau_{+}
= \tau_{+} - f_{-} (\tau_{+} - \tau_{-})
\]
or
\[
f_{-}=\frac{\tau_{+}-\tau_{Observed}}{\tau_{+} - \tau_{-}}
\]
Where $\tau_{Observed}$ is the measured value for the mean muon decay time in
the muon neutrino sample.

In general detectors only sample the muon decay rate in a time window following
the interaction so that there is a correction to this expression for
truncation of the interval.  For a data sample restricted to the time
range $t_{1}<t<t_{2}$, $\tau_{\pm}$ in  the expression above is modified to:
\[
\tau_{\pm} \rightarrow \tau_{\pm}
\frac{e^{(t2-t1)/\tau_{\pm}}(1+\frac{t_1}{\tau_{\pm}})
-(1+\frac{t_2}{\tau_{\pm}})}
{e^{(t2-t1)/\tau_{\pm}}-1}
\]

A cleaner result might be obtained by fitting
the observed time distribution to extract $f_{-}$ (and confirm the values
of $\tau_{+}$ and $\tau_{-}$).
A number of consistency
checks are possible.  The fraction of decays attributable to $\mu^{-}$
decreases more rapidly than for $\mu^{+}$ so one may get a more accurate
measurement by using a delayed sample of decays.  All such temporal subsamples
must yield a consistent value for $f_{-}$.

With sufficient statistics
this method can be exploited in bins of neutrino energy or flight
distance which are the relevant observables for the oscillation hypothesis.
Vacuum oscillations should show no difference in the $\mu^{+}$ to
$\mu^{-}$ fraction so differences would be a clear indication of new
physics.

\section{Complications}
The value of $f_{-}$ is a measurement of the $\mu^{-}$ fraction of the muon
decay sample.  It is at best
an indirect measurement of the 
$\nu_{\mu}$ to $\overline{\nu_{\mu}}$ flux ratio.  The cross sections for
interaction of these two neutrino types are quite different so the observed
value of $f_{-}$ must be corrected.

The triggering efficiency and reconstruction efficiency for $\nu_{\mu}$ and
$\overline{\nu_{\mu}}$ induced reactions may be different and must be
corrected.

Muon polarization effects may make the detection efficiency for the two muon
charges different.  Some accounting for a lower efficiency for observing a
signal from muon capture may be needed.

Subthreshold pion decays that give rise to the decay sequence
$\pi \rightarrow \mu \rightarrow e$ may also populate the post
interaction time distribution.  Subthreshold pions are additional tracks
in the initial neutrino interaction that escape detection but will
subsequently decay.  These can be studied in several ways.  The muon
decay time distribution observed in $\nu_{e}$ interactions can be subtracted
from that observed in muon type reactions.
Events with multiple muon decays can be studied to understand the rate
for which subthreshold pion decays occur.  With sufficiently high
detection efficiency the problem could be eliminated by removing events
in which more than one muon decay is observed.

A spatial cut might be possible in that muon decays occurring near the primary
interaction vertex are removed since this is where such subthreshold pion
decays would be found.

\section{Conclusions}
The atmospheric neutrino anomaly has been firmly established.
More information about the nature of the interactions is necessary
to fully understand the physical mechanism responsible for the effect.
By exploiting lifetime differences in muon decay one has access to the
$\nu_{\mu}$ and $\overline{\nu_{\mu}}$ fractions of events.
A sufficiently large sample would permit the study of the $\nu_{\mu}$ and
$\overline{\nu_{\mu}}$ content as a function of energy and distance.
The absence of any variation of this fraction with flight path
would strengthen the case for neutrino oscillations.  A variation
would point to some new physics.

\section*{Acknowledgements}
I would like to thank my BaBar colleagues for discussions about CP and CPT
violating observables.

\end{document}